\documentstyle[prb,multicol,aps]{revtex}

\begin{document}

\draft 
\tightenlines
\widetext

\title{Influence of electron correlations on
ground-state properties of III-V semiconductors}

\author{Simon Kalvoda, Beate Paulus, Peter Fulde} 
\address{Max-Planck-Institut f\"ur
Physik komplexer Systeme, Bayreuther Stra\ss e 40, D-01187 Dresden, Germany}
\author{Hermann Stoll} \address{Institut f\"ur Theoretische Chemie,
Universit\"at Stuttgart, D-70550 Stuttgart, Germany}

\maketitle

\begin{abstract}
Lattice constants and bulk moduli of eleven cubic III-V semiconductors
are calculated using an {\it ab initio} scheme.
Correlation contributions of the valence electrons, in particular,
are determined using increments
for localized bonds and for pairs and triples of such bonds;
individual increments, in turn, are evaluated using 
the coupled cluster approach with single and
double excitations.
Core-valence correlation is taken into account
by means of a core polarization potential. Combining the results at the
correlated level 
with corresponding Hartree-Fock data,  we obtain
lattice constants which agree with experiment within an average error
of $-$0.2\%; bulk moduli are accurate to  +4\%.
We discuss in detail the influence of the various correlation contributions 
on lattice constants and bulk moduli.
\end{abstract} 

\begin{multicols}{2}
\narrowtext

\section{Introduction}
Ground state properties of solids are most often
calculated within the framework of density-functional theory (DFT),
where the 
exchange and correlation contributions are determined within the 
local-density approximation (LDA) \cite{jones89}. 
These methods yield good 
results for lattice
constants and bulk moduli but do not provide
for many-body wave functions. Both, the influence of exchange and correlation 
are covered in
an implicit way, thus, a systematic improvement towards the exact 
results appears to be
difficult. \\
Hartree-Fock
self-consistent-field (HF-SCF) calculations for solids \cite{pisani88},
while lacking electron correlations as per its definition, have the
merit of treating non-local exchange exactly. They supply a good
starting point for subsequent correlation calculations
provided correlations are not too strong.
Infinite systems 
require the use of size-consistent approximations for electron correlations.
In Ref. \onlinecite{paulus95} an approach based on cumulants was 
discussed which ensures size-consistency at any level
of approximation. We use it here in order to work with local
operators, i.e., local one- and two-particle excitations.
These are best suited for describing the local correlation hole which the
electrons carry with them. The matrix elements which appear in the 
calculation of the ground-state energy are evaluated using clusters instead
of solids thereby
applying an incremental method\cite{stoll92}.
The same method has previously been used to calculate ground-state
properties of elementary semiconductors\cite{paulus95} as well as 
the cohesive energy
of cubic III-V compounds\cite{paulus96}.\\
In the present paper we apply this method to eleven III-V
compounds in zinc-blende structure and investigate the 
influence of electronic correlations on
lattice constants and bulk moduli. In addition to the 
valence correlations the core polarization, which may have large
influence on the bond lengths of molecules\cite{mueller84}, is taken into account
for the solid in a systematic way.

\section{Hartree-Fock calculation}
Before discussing the influence of electron correlation effects in
solids, reliable HF-SCF calculations are a necessary prerequisite.
We performed HF ground-state calculations for the III-V compounds 
using the program package
{\sc Crystal92} \cite{crystal92}. 
The same basis sets and pseudopotentials as in Ref.
\onlinecite{paulus96} are used\cite{korrektur}.\\
The HF lattice constants have been determined from the minima of the
HF ground-state energy curves in the following way. We varied the lattice constants in steps
of 1\% of the experimental values and calculated six points of
the potential curves in each case. The minima were evaluated from a
quadratic fit to these points. A narrower spacing and higher order fits 
were tested
but yielded only a marginal change in the position of the minimum 
and the curvature.
The results for the HF lattice constants are listed in Table I.\\
The bulk moduli, which describe
the response of the solid to a homogeneous pressure,
have been evaluated at the experimental lattice constants\cite{landoltiii17a}.
The HF values for the bulk moduli are listed in Table II.

\section{Correlation effects}

\subsection{Core polarization}

The description of core electrons by means of pseudopotentials
implies a neglect of static and dynamic core polarization.
The first part is an SCF effect, while the
latter part is related to core-valence correlation.
Especially if closed $d$ shells are described by
pseudopotentials, the influence of the core polarization on lattice constants 
and bulk moduli is significant although the influence on cohesive
energies is negligible. The computational effort for an explicit 
treatment of the core
electrons and their correlations would be very high; thus, we simulate these
effects in our calculations by means of a core polarization potential (CPP)
\cite{mueller84}. It describes the dipole interaction between
the core and the valence electrons:
\begin{equation}
V_{\rm CPP}=-\sum_{\lambda}\frac{1}{2} \alpha_{\lambda}
\vec{f}_{\lambda}^2 ;
\end{equation}
here $\lambda$ numbers different cores, $\alpha_{\lambda}$
is the corresponding dipole polarizability, and $\vec{f}_{\lambda}$ 
is the field at
the site of the core generated by valence electrons and surrounding cores:
\begin{equation}
\vec{f}_{\lambda}=
\sum_i^{N_{\rm V}} 
\frac{\vec{r}_{\lambda i}}{r_{\lambda i}^3}
\left( 1-e^{-\delta_{\lambda}r_{\lambda i}^2} \right) - 
\sum_{\mu (\neq \lambda)} 
\frac{\vec{r}_{\lambda \mu}}{r_{\lambda \mu}^3}
\left( 1-e^{-\delta_{\lambda}r_{\lambda \mu}^2} \right) ;
\end{equation}
$\vec{r}_{\lambda i}$ is the distance between valence electron
$i$ and core $\lambda$, $N_{\rm V}$ is the number of valence
electrons. The cut-off parameter $\delta_{\lambda}$ is necessary in order 
to remove 
the singularity of the dipole interaction at $r_{\lambda i}=0$.
We took the parameters $\alpha_{\lambda}$ and $\delta_{\lambda}$ 
from Ref.\ \onlinecite{igelmann88}, where the CPP was adjusted in atomic
calculations to the spectra of single-valence-electron ions.\\
In order to obtain information on the influence of CPP's on solid-state 
properties,
we choose a finite fragment of the zinc-blende structure --
a $X_4Y_4{\rm H}_{18}$ cluster -- where the
dangling bonds are saturated with hydrogen atoms.
The $X$---H and the $Y$---H distances,
respectively, are optimized in CCSD calculations for $XY{\rm H}_6$
clusters \cite{paulus96}.
For hydrogen we choose Dunning's  double-zeta basis\cite{dunning89}
without the  $p$ polarization function.  For the other elements, large-core
pseudopotentials in conjunction with corresponding 
$(4s4p1d)/[3s3p1d]$ valence basis sets are used\cite{bergner93,paulus96}.\\
We first want to check the local character of the CPP.
Test calculations for GaAs are performed in the following way.
The CPP for Ga is only provided at the innermost Ga atom and the
energy gain due to the CPP, $\Delta E_{\rm CPP}(\rm Ga)$, is calculated 
at the HF-SCF level with
the program package {\sc Molpro94}\cite{molpro94}.
In the next step we provide the CPP only at the innermost As atom,
and $\Delta E_{\rm CPP}(\rm As)$ is determined.
Providing the CPP's at the inner Ga and the inner As atoms simultaneously,
we finally calculate $\Delta E_{\rm CPP}(\rm Ga,As)$, and we find 
that the non-additive
part $\Delta E_{\rm CPP}(\rm Ga,As) - (\Delta E_{\rm CPP}(\rm As)+
\Delta E_{\rm CPP}(\rm Ga))$ is negligibly small.\\
With this knowledge in mind, we determine an approximation for the 
core-polarization energy
of the solid per unit cell by calculating
$\Delta E_{\rm CPP}(X,Y)$ in the $X_4Y_4{\rm H}_{18}$ cluster as
described above.
Varying all $X$---$Y$ bonds in the
$X_4Y_4{\rm H}_{18}$ cluster,
we obtain the dependence of $\Delta E_{\rm CPP}(X,Y)$ on the 
lattice constant $a$.
The results at this level
are listed in Table I for the lattice constants and
in Table II for the bulk moduli; they include core-valence correlation 
(through the use of the CPP) but do not account for the
coupling between core-valence and valence correlation, which is dealt 
with in the next subsection.

\subsection{Valence correlations}

As pointed out in the introduction, the matrix elements which appear
in the calculations of the ground-state energy are obtained from
replacing the solid by  
sufficiently large clusters. With these matrix elements the correlation 
contribution to the ground-state energy can be partitioned into a sequence 
of increments\cite{paulus95,stoll92} which are successively evaluated.
The procedure and the computational details for the
III-V compounds are described in Ref. \onlinecite{paulus96}.
In order to account for the valence/core-valence correlation-energy 
coupling, we included CPP's on every
$X$- and $Y$-atom and used the HF-SCF ground state with CPP as 
reference state for the
correlation calculation.
For the evaluation of the largest increments (one-bond and up to next-nearest-neighbour two-bond)
the basis set was extended to $(4s4p2d1f)$\cite{paulus96}.
Applying this method we determined the 
correlation energies as function of the lattice constant. 
The calculated lattice constants and bulk moduli are listed in Table I
and Table II.

\section{Results and Discussion}

We have calculated the lattice constants and the bulk moduli for 
eleven cubic III-V semiconductors, at different theoretical levels.
A HF treatment overestimates the lattice constants by up to 2\% ,
with the errors increasing for systems with heavier atoms. Apparently, 
it is due to the lack of
electron correlations which describe the instantaneous motion
of the electrons, that the structural minimum is found at larger 
volumes as compared to experiment.
This is opposite to what is usually found for molecules composed of first- 
and second-row atoms in 
quantum-chemical calculations. There, the lack of non-dynamical correlation 
(left-right correlation),
yields an artificial upwards shift of the potential-energy curve for large 
$R$ and consequently leads to too short
SCF bond lengths. Of course, this effect is also present in the solids 
considered in this paper, but apparently
is overridden by dynamical correlation effects (cf.\ below).
However, quite in common with quantum-chemical experience for molecular 
force constants,
the bulk moduli are overestimated
by up to 30\%, at the  HF level. The materials seem to be harder 
(large bulk modulus)
because the instantaneous answer of the electrons to pressure
is neglected.\\
In the next step we take into account the core polarization. As described 
in Section III.A,
the core polarization is neglected when using
pseudopotentials. Pictorially
speaking, the instantaneous deformation of the spherical cores in the 
field of the 
valence electrons is omitted. Modelling this effect with
a core polarization potential, we achieve a 
significant reduction of the lattice constants (by up to -2\%).
The calculated values for the bulk moduli decrease by up to 15\% 
depending on the materials (heavier atoms are easier to polarize).\\
The overall effect of the valence correlations determined with the method 
of local
increments leads to a reduction of about 1\% in the lattice constant and
about 10\% in the bulk modulus.\\
To get a deeper understanding of the nature of the interplay of the 
various correlation effects, we
performed calculations for GaAs with different basis sets (see Table III).
The interatomic (non-dynamical) part of the valence 
correlation energy is determined using a minimal
$(4s4p)/[1s1p]$ basis set. It enlarges the lattice constant by more than 2\%,
because in the many-body wave function only excitations
into anti-bonding orbitals are possible. 
Intra-atomic (dynamical) correlations (calculated with a larger than minimal 
basis) decrease the lattice constant because more low lying excitations
are mixed into the interatomic many-body wave function
within the enlarged configuration space. The electrons
can avoid each other much better. Increasing the basis further
(to $4s4p2d1f)$\cite{paulus96}, the intraatomic correlations are
increased and lattice constants and bulk moduli are decreased.\\
A comparison between valence-correlation effects
with and without CPP has also been  made for GaAs (see Table III).
This way, the coupling between valence and core-valence correlation 
can be investigated.
While the interatomic part of the valence-correlation energy is 
virtually unchanged
when including the CPP, the
intraatomic one is reduced in the calculation with CPP.
The reason for this effect, however, is not so much a reduction of 
dynamical valence correlation
but rather a reduction of dynamical core-polarization when angular 
correlation around the atoms is introduced 
into the many-body wave function: angular correlation leads to a 
preference for valence electrons occupying 
opposite positions with respect to an atomic core so that their 
polarizing effect is reduced as compared
to an independent-particle (uncorrelated) description.
The major part of this effect comes from
$s$, $p$ and $d$ functions, the additional bond-length reduction due
to $f$ functions is nearly the same in the calculation with 
and without CPP.\\
For the lattice constants we reach a very good agreement with 
experiment. All calculated results are slightly too small (by -0.2\% 
on the average),
but they are within the error bars given by the
pseudopotentials and the CPP's ($\approx$ 0.03 \AA ).
We have checked the quality of the  chosen basis set for
the Ga---As bond length in GaAsH$_6$. There, we  see nearly
no change ($\leq$0.1\%) when extending the basis to an even larger
one ($6s6p3d2f1g$). The average error of the calculated
bulk moduli is about +4\% .\\
For comparison, we have also listed in Table I and Table II results
from the literature which have been obtained with LDA. Lattice
constants are underestimated within the LDA treatment by up to 2.8\%.
Bulk moduli mostly agree well with experiment. 
    
\section{Conclusion}

We have determined the lattice constants and the bulk moduli of 
eleven cubic III-V semiconductors. At the HF-SCF level,
the lattice constants are overestimated by up to 2\%,
the bulk moduli by up to 30\%.
A significant reduction towards the experimental values 
is obtained when applying  core polarization 
potentials. Valence correlation has been calculated 
at the CCSD level, using the method of local increments.
For the lattice constants two 
opposite trends are seen: interatomic correlations enlarge
the lattice constant, while intraatomic ones decrease it. The
overall trend is a reduction of the lattice constants,
and the final results agree very well with experiment
(average error $\approx$-0.2\%). Valence correlations
substantially decrease the bulk moduli, and again the agreement with experiment
is remarkable good (average error $\approx$+4\%).
Due to the availability of many-body wave functions in our scheme, the influence of correlations
on the lattice constants and the bulk moduli can be discussed in detail.

\end{multicols}
\widetext

\begin{table} 
\caption{Lattice constants (in  \AA\ )
at different theoretical levels (cf.\ text); deviations
from experimental values are given in parentheses;
the experimental values measured at room temperature
\protect{\cite{landoltiii17a}} have been extrapolated to zero Kelvin.}
\begin{tabular}{c||cc|cc|cc|cc|c}
&\multicolumn{2}{c|}{$a_{\rm HF}$}&\multicolumn{2}{c|}{$a_{\rm HF+CPP}$}&
\multicolumn{2}{c|}{$a_{\rm HF+CPP+corr}$}&\multicolumn{2}{c|}
{$a_{\rm LDA}$}&$a_{\rm exp}$\\
\hline
BP&4.5836&
(+ 1.0\%)&4.5679& (+0.7\%)&4.5322&(-0.1\%)&4.474
\protect{\cite{rodriguez95}}&(-1.4\%)&4.5373\\
BAs&4.8276&(+1.1\%)&4.8011&(+0.5\%)&4.7644&(-0.3\%)&4.777
\protect{\cite{wentzcovitch86}}&($\pm$ 0.0\%)&4.777\\ 
AlP&5.5348&(+1.3\%)&5.5016&(+0.7\%)&5.4398&(-0.5\%)
&5.421\protect{\cite{yeh92}}&(-0.8\%)&5.4672\\
AlAs&5.7405&(+1.5\%)&5.6944&(+0.6\%)&5.6428&(-0.3\%)&5.620
\protect{\cite{yeh92}}&(-0.7\%)&5.6600\\
AlSb&6.2606&(+2.1\%)&6.1954&(+1.0\%)&6.1307&($\pm$0.0\%)&---&---&6.1355\\
GaP&5.5298&(+1.5\%)&5.4636&(+0.3\%)&5.4274&(-0.3\%)
&5.3581\protect{\cite{agrawal95}}&(-1.7\%)&5.4459\\
GaAs&5.7546&(+1.9\%)&5.6730&(+0.4\%)&5.6472&($\pm$0.0\%)&5.508
\protect{\cite{agrawal95}}&(-2.5\%)&5.6485\\
GaSb&6.2120&(+2.2\%)&6.1034&(+0.4\%)&6.0740&(-0.1\%)&5.939
\protect{\cite{agrawal95}}&(-2.3\%)&6.0806\\
InP&5.9527&(+1.5\%)&5.8559&(-0.2\%)&5.8296&(-0.6\%)&5.7021
\protect{\cite{vancamp90}}&(-2.8\%)&5.8666\\
InAs&6.1524&(+1.6\%)&6.0521&($\pm$0.0\%)&6.0437&(-0.2\%)&5.9019
\protect{\cite{vancamp90}}&(-2.5\%)&6.0542\\
InSb&6.5925&(+1.9\%)&6.4717&($\pm$0.0\%)&6.4615&(-0.2\%)&6.3406
\protect{\cite{vancamp90}}&(-2.0\%)&6.4719\\
\end{tabular} 
\end{table}

\begin{table}
\caption{Bulk moduli (in Mbar)
at different theoretical levels (cf.\ text); deviations
from experimental values \protect{\cite{landoltiii17a}}
are given in parentheses.}
\begin{tabular}{c||cc|cc|cc|cc|c}
&\multicolumn{2}{c|}{$B_{\rm HF}$}&\multicolumn{2}{c|}{$B_{\rm HF+CPP}$}&
\multicolumn{2}{c|}{$B_{\rm HF+CPP+corr}$}&
\multicolumn{2}{c|}{$B_{\rm LDA}$}&$B_{\rm exp}$\\
\hline
BP&1.95&(+12\%)&1.92& (+11\%)&1.87&(+8\%)&1.73
\protect{\cite{rodriguez95}}&($\pm$ 0\%)&1.73\\
BAs&1.62&---&1.53&---&1.51&---&1.45
\protect{\cite{wentzcovitch86}}&---&---\\
AlP&1.01&(+17\%)&0.98&(+14\%)&0.92&(+7\%)&---&---&0.86\\
AlAs&0.90&(+17\%)&0.86&(+12\%)&0.82&(+7\%)&---&---&0.82\\
AlSb&0.65&(+12\%)&0.61&(+5\%)&0.57&(-1\%)&---&---&0.58\\
GaP&1.10&(+21\%)&0.98&(+8\%)&0.93&(+2\%)&0.9783
\protect{\cite{agrawal95}}&(+7\%)&0.91\\
GaAs&0.87&(+13\%)&0.82&(+6\%)&0.77&(0\%)&0.771
\protect{\cite{agrawal95}}&($\pm$ 0\%)&0.77\\
GaSb&0.63&(+12\%)&0.63&(+12\%)&0.59&(+6\%)&0.7994
\protect{\cite{agrawal95}}&(+43\%)&0.56\\
InP&0.79&(+10\%)&0.71&(-1\%)&0.66&(-8\%)&0.7614
\protect{\cite{vancamp90}}&(+6\%)&0.72\\
InAs&0.74&(+28\%)&0.69&(+19\%)&0.64&(+10\%)&0.6190
\protect{\cite{vancamp90}}&(+7\%)&0.58\\
InSb&0.58&(+26\%)&0.52&(+11\%)&0.48&(+4\%)&0.4774
\protect{\cite{vancamp90}}&(+4\%)&0.46\\
\end{tabular}
\end{table}

\begin{table}
\caption{Influence of correlations on lattice constants (in \AA) 
of GaAs, employing different basis sets
(cf. text); deviations
from experimental values 
are given in parentheses.}
\begin{tabular}{c||cc|cc|cc|cc}
&\multicolumn{2}{c|}{HF}&\multicolumn{2}{c|}{CCSD, min. basis}&
\multicolumn{2}{c|}{CCSD,$3s3p1d$}&\multicolumn{2}{c}{CCSD, $4s4p2d1f$}\\
\hline
$a$(with CPP)&5.6730&(+0.4\%)&5.7910&(+2.5\%)&
5.7084&(+1.1\%)&5.6472&($\pm$0.0\%)\\
\hline
$a$(without CPP)
&5.7546&(+1.9\%)&5.8810&(+4.1\%)&5.7645&(+2.1\%)&5.7015&(+0.9\%)
\end{tabular}
\end{table}


\begin{thebibliography}{10}
\bibitem{korrektur} The HF basis set of BP had to be changed 
with respect to the one of Ref. 5. The very diffuse $d$-projector in
the boron pseudopotentials is now neglected in the basis optimization.
We optimized a new HF basis $(4s4p1d)/[3s3p1d]$ for BP 
(two outer uncontracted $s$ exponents for B: 0.58,0.20;
two outer uncontracted $p$ exponents for B: 0.53,0.32;
$d$ exponent for B: 0.45; $d$ exponent for P: 0.55 ).
\end{thebibliography}
\end{document}